\documentstyle[prl,aps,amssymb,graphicx]{revtex}

\newcommand{\om}{\omega_m}

\newcommand{{ \vn }}{\vec{n}}

\newcommand{\beq}{\begin{equation}}
\newcommand{\eneq}{\end{equation}}

\newcommand{\met }{\frac{1}{2}}
                                              \begin{document}

           \twocolumn[
           \hsize\textwidth\columnwidth\hsize\csname@twocolumnfalse\endcsname

           \title{Spin fractionalization of an  even number of electrons 
              in  a  quantum dot }
            \author {  Domenico Giuliano$^{1,2}$  and
             Arturo Tagliacozzo$^{1,3}$}
        \address{ $^{1}$Istituto Nazionale di Fisica della Materia (INFM),
         Unit\'a di Napoli\\
         $^{2}$ Department of Physics, Stanford University,
         Stanford, California 94305\\
         $^{3}$Institut f\"ur Nanotechnologie (INT), Forschungszentrum,
         Postfach 3640,D-76021 Karlsruhe, Germany    \\ 
         and     Dipartimento di Scienze Fisiche Universit\`a di Napoli 
         "Federico II ",\\
         Mostra d' Oltremare Pad. 19, I-80125 Napoli, Italy}
         \maketitle

\begin{abstract}
Kondo resonant conductance can occur in a  
  quantum dot  with an even number of 
electrons $N$ in the Coulomb blockade regime,
with bottom and top contacts attached to it in a pillar configuration,
 in a  magnetic field $B_{\perp} = B^* $ along the axis. 
The field $B^*$ is  tuned at the  crossing  of levels with 
different total spins and 
  the spin degeneracy of the dot energies  should be lifted. 
  The case of the  degeneracy  of the singlet-triplet states 
 in a dot with $N=2$ is 
analyzed in detail.  Coupling to the contacts is antiferromagnetic 
due to a spin selection rule, so that total spin on the dot is $S=1/2$ 
in the Kondo state. 
\end{abstract}

\pacs{ 
\hspace{1.9cm} 
PACS numbers:{83.30.Vw,79.60.Jv,72.15.Qm}}
]

\narrowtext

Quantum dots ($QD$) are remarkable because of Coulomb oscillations
\cite{kouwenhoven}. 
At very low temperature, linear conductance (at vanishing bias
$V_{sd}$)  is zero, except for peaks at discrete 
values of the gate voltage $V_g$, when  it is energetically favourable to 
add one extra electron to the dot. Therefore $V_g$ controls the 
number of  particles on the dot in the 
Coulomb Blockade $(CB)$ regime: its charge   is quantized 
 and so is the spin. Dots with an even (odd) number of electrons have 
integer ( half integer ) total spin.

However,by lowering the temperature further below some value $T_K$, 
finite conductance can be  found experimentally at zero voltage
\cite{goldhaber,cronenwett,schmid}, relatively 
insensitive to small changes of $V_g$ within the CB regime,
 as a mesoscopic realization of the physics of the 
Kondo effect \cite{glazman,ng,meir}. The zero voltage anomaly in the 
conductance is only seen when population is odd \cite{cronenwett}.
While the charge remains quantized, the spin quantization 
of the correlated state between the dot and its contacts  is changed, 
due to the formation of a Kondo singlet involving the odd electron.

A weak magnetic field parallel to the dot plane, $B_{\parallel}$, splits 
the peaks in the differential conductance into two, 
located at  $V_{sd} =\pm g^* \mu_B B_{\parallel}$,
 (where $g^*$
is the effective giromagnetic factor for electrons in this geometry and
 $\mu_B = e\hbar/2 m c$ is the  Bohr magneton),until  eventually they
 disappear \cite{cronenwett,schmid}.

On the contrary, a magnetic field $B_{\perp}$, orthogonal  to the dot area, 
 is believed to be much more destructive, due to 
  orbital changes of the electron 
states\cite{xue}, in addition to the lifting of spin degeneracy , that 
prevents the formation of the Kondo correlated state.  

We suggest here the possibility that, due to accidental
crossing of the lowest non degenerate
many body energy levels of the $QD$  at special values of 
$B_{\perp} =B^*$, 
orthogonal to the dot connected to the leads in a pillar structure,
 the Kondo resonance can  still be present, 
irrespective of the parity of the electron number $N$
\cite{pustilnik,nota0}.
 In particular 
we discuss here the case of $N=2$ and show that under 
suitable conditions a correlated  state sets in, which is characterized
by fractionalization of the total spin.

Indeed  a $QD$ at high $B_{\perp}$ (in the $z- direction$)
 displays a rich level structure that can be monitored attaching 
bottom and top contacts to it in a pillar structure via linear 
and non linear transport\cite{kouwenhoven2,tarucha,faini,jouault}.
Assuming the $QD$ to be effectively two-dimensional  and confined by a 
parabolic potential, its states in isolation can be labelled by the
 total spin  $S$ 
and the $z-$component of the total angular momentum $M$  and
spin  $S_z$.  In increasing  $B_{\perp}$,
 many level crossings can occur for an $N$ particle $QD$  between 
states of increasing $M$, as well as states with higher total spin $S$.
  
\begin{figure}
\includegraphics*[width=\linewidth]{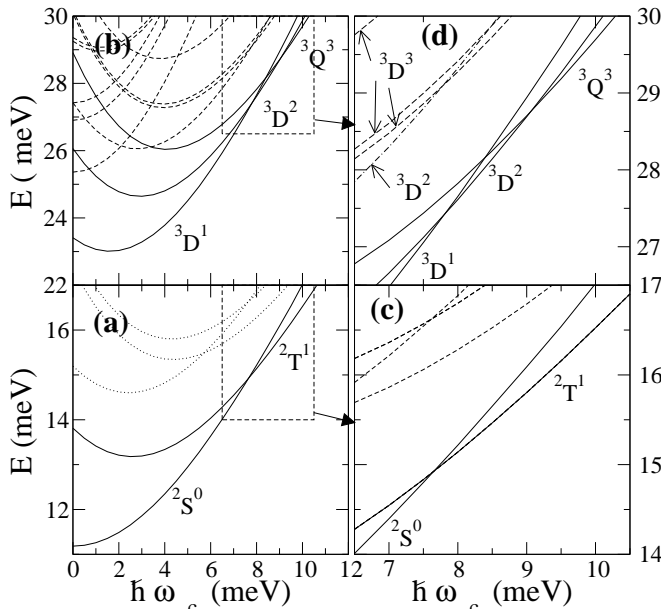}
\caption[]{ Calculated energies of a circular quantum dot {\sl vs}
 magnetic field $B_{\perp}$ (expressed  in cyclotron energies 
$\hbar \omega_c$) for $N=2$ {\bf (a),(c)} 
and $N=3$ {\bf (b),(d)}. Here the level spacing 
of the confining potential is $\hbar \omega_d = 
4meV$ and the strength of the Coulomb interaction  is $E_c= 3meV$
 (notations can be found in the text). { \bf (c),(d)}
magnify the marked area of  { \bf (a),(b)}, respectively \cite{jouault}}.  
\label{fig}
\end{figure}

It is well known that the  ground state (GS)  of an isolated dot  
  with $N=2$ electrons changes from singlet($^2S :  S=0, M=0 $) to 
triplet ($ ^2T_{S_z}: S=1, M=1 $) at a value $B_{\perp} =B^* $,
due to crossing of levels\cite{wagner}.  Correspondingly, exact 
diagonalization shows that  the $N=3$ electron dot  is in a doublet state
($ ^3D_{S_z}: S=1/2, M=1 $ or $2$ depending on the field value)
(see fig.\ref{fig})
\cite{jouault}.  As shown quantitatively in the following, the dot 
parameters can be chosen in such a way that  $B_{\perp}$ is high 
enough  to lift  spin degeneracy. 
We take $V_g$  tuned in the $CB$ region between
 $N=2$ and $N=3$ electrons on the dot, and limit our discussion to four 
states : 1) dot with $N=1$ in  $^1D$, 2-3) dot with $N=2$ in one of the 
two degenerate states $^2S_0,$
and $^2T_1$, 4) dot with $N=3$ in $^3D_{1/2}$.  Taking only these states 
allows the  mapping of the problem onto the non degenerate Anderson model 
with the following levels: 1) empty 
level of zero energy, 2-3) singly occupied level of 
energy $\epsilon_d$ (doubly degenerate),  
 4) doubly occupied level  of energy $2\epsilon _d +U $.
Here $U$ is very large (see below)  and, if levels 2-3 and 4 are  symmetric 
with respect to the  chemical potential of the contacts
$\mu$, what can be  controlled by $V_g$, average occupancy of 
levels 2-3  is $n_d =1$.  This corresponds to $N=2$ electrons on the dot. 

The key  observation is the following spin selection rule for the isolated 
dot\cite{weinmann} :
 because  the $N=3$ GS  for the dot 
only has $S_z=1/2$, hybridization 
of the dot with the contacts  can only occur by virtual occupancy of the 
dot with  an extra spin up electron  when the dot is in the $N=2$ 
singlet state $^2S$ or with an extra spin down electron when the dot is in the 
$N=2$ triplet state with $S_z =1$ ($^2T_1$ state). Although spin degeneracy
 of the dot is assumed to be lifted, the physics is the same as that of 
the  single channel Kondo model  with effective 
spin $S^{eff}=1/2$, {\sl in the absence of magnetic field}.
  Here is $S^{eff}_z=S_z -1/2$
 and coupling with the 
delocalized electrons is antiferromagnetic due to the spin selection 
rule mentioned above. Provided hybridization to the contacts 
$\Delta $
is large enough and temperature is below the Kondo 
temperature  $T_K $,
 the system 
flows to the strongly correlated fixed point. Formation of the Kondo singlet 
corresponds in this case  to an average spin on the dot frozen  
at $S_z=\met$, that is to fractionalization of the total spin of the dot,
together with a strong zero bias anomaly  in the conductance.
 
In this letter we discuss the properties of the GS, borrowing from the 
known results of   the  non degenerate Anderson model\cite{tsvelick}.
In particular, we discuss the spin fractionalization, 
the practical realizability of the system 
and the robustness of the GS  versus
small changes of the magnetic field $B_{\perp}$  with respect to the 
value $B^*$ that makes the states$| S\rangle$ and  $|T\rangle$ degenerate in 
energy.


We denote   the two dot
degenerate  energies  by $^2E_i$ 
($i = 1:$ triplet state, $i= 2:$  singlet state) and take the 
chemical potential  $\mu$ of the left and right
bulk contacts within the  CB  valley of the conductance at $N=2$ 
 ($ \mu _N < \mu <\mu _{N+1} \equiv
\: ^{N+1}\!E -^N\!E_i )$. Electrons in the contacts  are non interacting 
and have  
an energy dispersion $\epsilon_{k\sigma}$ involving just one single Landau 
level and 
 linearized  around  $\mu$ with 
Fermi velocity $v_{F\sigma}$. The wavevector $k$ is in the $z-$ direction 
and the two contacts are assumed to be equal
($ H_{L/R}= \sum_{k\sigma} \epsilon_{k , \sigma}
b^{\dagger}_{L/R\;k\sigma} b_{L/R\;k\sigma}$)
 and three dimensional enough, in order to have both  spin polarizations
at $\mu$. 

The dot Hamiltonian $H_d$, restricted to the Fock space with $N
=1,2,3$ electrons
 can be represented in terms of the occupation numbers  $n_{i}$
 for  the
degenerate  levels $^2E_i$:
\beq
H_d = \mu _2\sum_i n_i + U n_1 n_2
\label{hdot}
\eneq
where $ U=\mu _{3} -2\mu _2 $ is the charging energy for adding the
third electron to the dot. Here  is $n_{i}= c^{\dagger}_{i} 
c_{i}$, the vacuum of the $c$ operators is the dot state with 
one electron and the $c^\dagger$  
create the two degenerate states by acting on it.

The model Hamiltonian  is $H = H_L + H_R + H_d +H_t $, where 
$H_t$ is the hybridization term acting only at $z=0$.
 According to the spin selection 
rule stated above, it takes the form:

\[
H_t = \frac{1}{\sqrt{2}} \sum _{k} \left \{ \Gamma^*
(b^{\dagger}_{L\;k\uparrow}+ b^{\dagger}_{R\;k\uparrow}) c_{1}+
\Gamma c_{2}^\dagger (b_{L\;k\downarrow}+
 b_{R\;k\downarrow}) \right \} + h.c.
\]

The tunneling amplitudes 
$\Gamma_{L/ R\;k\sigma i}$, which depend on the actual features 
of the device, have been taken all equal for simplicity.

We start integrating out  
 the contacts fields  ($\hbar =1$ in the following):
\beq
{\cal{Z}}(\mu)  \propto
\int  \Pi_i \left ( D c_i D c^\dagger_i \right )  e^{  -  {\cal {A}}_d } \:
 e^{-\beta |\Gamma |^2 \sum_{\om, i } c^\dagger _i K(i\om )c_i}
\label{part}
 \eneq
with
 
\[
{\cal {A}}_d = \int_0^{ \beta } d \tau
\left \{ \sum_{i} \left [ c_{i}^{\dagger}
\frac{\partial}{\partial \tau } c_{i}+
(\mu _2-\mu)\: c^{\dagger}_{i}
c_{i} \right ] + U c^{\dagger}_1 c_1 c^{\dagger}_2 c_2
\right \}\: \nonumber
\]

and 
$ K(\om ) = \sum _k ( i\om +vk)^{-1} $ is taken independent of $\sigma$ for 
simplicity.
In the limit of large U we have:
\[
\exp [ \int_0^\beta d\tau [Un_1n_2-\tilde{\mu}(n_1+n_2)] ] 
\]

\[
= e^{\frac{\beta}{4 U}\tilde{\mu}^2}\; \delta (n_1+n_2 
- 2\tilde{\mu}/U)
\: \cdot e^{-\frac{U}{4}\int_0^\beta \;d\tau (n_1-n_2)^2} \:\:,
\]

where the delta function implements the constraint of single site 
occupancy  in the symmetric case,  $\tilde{\mu} =\mu - \mu_2 = U/2 $.

We introduce $X( \tau )$ as  an auxiliary boson field coupled to the 
differential
occupancy of the two degenerate states by means of an Hubbard-
Stratonovitch decoupling:

\[
e^{ \frac{U}{4} \int_0^\beta d \tau ( n_1 - n_2 )^2 } =
\]

\[
\int D X e^{ - \frac{1}{4 U} \int_0^\beta d \tau ( X^2 ( \tau )
+ 2U ( n_1 - n_2 ) X ( \tau )) }.
\]

Then, the partition function of eq.(\ref{part})reads:

\[
{\cal{Z}}(\mu) \propto
\int D X e^{ - \frac{1}{4U} \int_0^\beta d \tau  X^2 ( \tau )}
\]

\begin{eqnarray}
\times
 \int   \Pi_i \left ( Dc_i Dc_i^\dagger 
 e^{ - \int_0^\beta d \tau d \tau ' c_i^\dagger ( \tau )
G^{-1}_{(0)}(\tau -\tau ')c_i (\tau ')} \right )  
\nonumber \\ \times \delta (n_1+n_2 -1) 
 \sum_{j=1,2} e^{  (-1)^{j}\int _0^\beta d \tau 
 [ n_j - \met ] X ( \tau )} \:\: ,
\label{part2}
\end{eqnarray}

where is $G^{-1}_{(0)}(i\om) = i\om +|\Gamma |^2 K(i\om)$.
We will take the saddle point solution of eq.\ref{part2}.
At saddle point (see below)  the constraint is satisfied in 
the average over the time $\beta$ because 
$ \langle n_i -\met \rangle \approx - (-1)^{i}  
\overline{X_{s.p.}(\tau )}/2U =0$ and
the integrals for $i=1,2$ decouple, giving a contribution to the
effective action for the field $X$ that, for $i =1$, is given by
$   {\int _0^1 d g  \int _0^\beta d \tau 
 [G_{(gX),1}(\tau,\tau ^+)  -\met ] X ( \tau )}$
with  
 $G^{-1}_{(gX),1}  = G^{-1}_{(0)}- g X$ ( $X\to -X$ for $1\to 2$)
\cite{hamann}.
This representation  allows to recognize  the doubly degenerate
GS of an effective spin  $S^{eff}=1/2$
with  $S^{eff}_z= (S_z -1/2)$, driven by the field $X(\tau)$ and
produced by virtual tunneling of electrons on and off the dot
at energy $\mu$.

 We now follow ref.\cite{hamann} quite
closely, to map eq.\ref{part2}  on a 1-dimensional Coulomb gas (CG).
We define  $\xi(\tau ) =  X (\tau )/\Delta$, where   
 $\Delta =2\pi N(0)\Gamma ^2  $ and  $N(0) = L/\pi  v_F$
($L$ is the size of each single  contact) is  the density
 of states at the Fermi energy.  
At low temperatures,  saddle point solutions of the resulting 
single particle 
effective action are  sequences  of instantons 
$ \xi_{\pm} (\tau, l ) = \pm \xi _0 
\tanh (( \tau - t_l) /\tau_0)$ ($t_l$ are the centers,
$l=1, 2, \ldots$)   corresponding to jumps
between
the two minima of the effective  potential,
$ V [ \xi ] = \Delta^2 / U \xi^2 - 2 \Delta / \pi [
\xi {\rm tg}^{-1} ( \xi ) - \frac{1}{2} \ln ( 1 + \xi^2 )$ \cite{hamann}, 
located at  $\xi_0 = \pm U/2\Delta$, that
  interact via a
logarithmic potential $\alpha ^2 \ln | (t_i - t_j )/\tau_0 |$.

The bare strength of the logarithmic interaction is $\alpha^2_b =
2(1- 4\Delta/U\pi^2)$.
The bare fugacity of the CG
 is \cite{yuval}
$Y_b = \tau_0 \exp (-\bar{{\cal {A}}})$ ($\bar{{\cal {A}}} 
\sim \tau_0 U$ is the action
of one  single blip). 
The scaling of the fugacity and the renormalization of the coupling
constant induced by processes of fusion of charges lead to the Renormalization
Group (RG) equations \cite{nie}:

\beq
 \frac{ d \bar{Y}}{ d \ln \tau_0 } = ( 1 - \frac{\alpha ^2}{2}) \bar {Y}
\; ; \frac{ d \alpha ^2}{ d \ln \tau _0 } =  - 2 Y^2 \alpha ^2
\label{rg1}
\eneq

 The flow is towards $ Y \to \infty $
and $\alpha ^2 \to 0 $ and the scaling invariant energy is 
$T_K = \tau_0^{-1} 
e^{1/ ( 1 - \frac{\alpha ^2}{2})}\sim (U\Delta)^{\met} 
e^{ -\pi^2 U/(4\Delta)}$. Condensation of instantons  in the
doubly degenerate  GS leads to the 
Kondo singlet   $ <S^{eff}> =0$ or $<S>= \met $ on the dot.

An heuristic  argument for the fractionalization of the total spin $S$ of the 
dot reads as follows. 

The dynamics of the field $\xi(\tau)$  of action  $\bar{{\cal {A}}}$
 can be mimicked  
by  a two level system hamiltonian $H_{2l}$ with hopping energy 
$\lambda \sim \met m_{eff} \left ( \frac{d\xi}{d\tau}\right )^2 
\sim \met \frac{\Delta^2}{U}\tau _0^2 (\xi_0/\tau_0)^2 \sim U/2$.
The role of the interaction is to project out higher energy 
components from the dot states. Denoting by $|\pm> $ the two eigenstates of 
 $H_{2l}$, the dynamics of the dot  is between states 
$\Pi^g|S>/{\cal{N}}_S^{\met }(g) $ and $\Pi^g|T>/{\cal{N}}_T^{\met }(g) $,
where $|S>$ and $|T>$ are the states localized in the two potential wells
 and ${\cal{N}}_{S/T} (g) $ are 
their  normalizations.   
$\Pi ^g = 1-g\Pi^-  $  is the Gutzwiller operator
partly eliminating  the component on the high energy eigenstate $|->$
$(g \in (0,1))$ \cite{vol}.

Now we state the connection with the CG picture 
given above.  The frequency
$\lambda / 2\pi  $ is obviously related to the average number of flips:
$  \beta \lambda /2\pi =<N>=Y\tanh Y$ and probability 
of having the system in states $|\pm>$ at temperature $\beta ^{-1}$ are 
\beq
P_- = \frac{e^{-4\pi Y\tanh Y}}{1+e^{-4\pi Y\tanh Y} }  \:\: ;\:\:
P_+ = \frac{1}{1+e^{-4\pi Y\tanh Y} } \cdot
\nonumber
\eneq
Equating these probabilities to the ones given by $P_{\pm} 
=\left | < \pm \left  | \Pi ^g \right  | S/T >\right |^2 /
{\cal{N}}_{S/T} (g) $ yields the correspondence $1-g \to e^{-2\pi Y\tanh Y}$.

The $z-$ component of the total spin of the projected singlet and 
triplet state are given by:
\begin{eqnarray}
\langle S_z \rangle _{S,g} = \frac{g^2}{2}\frac{1}{1+(1-g)^2} \:\: ; \:\:\:
 \langle S_z \rangle _{T,g} = 2 \frac{(1- g/2)^2}{1+(1-g)^2} \:\:\cdot
\nonumber
\end{eqnarray}

Because $Y$ scales to infinity, $g \to 1 $: the higher energy state 
completely decouples  and $<S_z>_{S/T,g} \to \met $. 

We  now give some estimates for a possible practical realization.

If the  level separation of the 
 parabolic confining potential of the dot is  $\hbar \omega_d =4 meV$,
the effective frequency in presence of $B_{\perp}$
is  $\omega_0 = \sqrt {\omega _d^2 + \omega _c^2/4 }$ 
($\omega_c =eB_{\perp}/m^*c$
 is the  cyclotron frequency  with $m^* =0.067 m_e $ for $GaAs$). The
Coulomb  interaction between electrons in the plane  is 
$E_c\times l_0/|\vec{r}-\vec{r}'|$ 
( where $E_c  = 3  meV$ and  $l_0^2 = \hbar/m^*\omega_0$).
   This choice of the parameters  gives a singlet-triplet
 (S-T) transition   for the dot 
with $N=2$ at $B_{\perp}=B^* \approx 4.5\; Tesla$,
close to  one of the Delft  experiments 
\cite{kouwenhoven2}. Fig.\ref{fig} shows the total energy of the dot 
 $vs$ magnetic field for $N=2$ and $N=3$. Zeeman spin splitting 
of the energies is 
{\sl not}
included in the picture  for clarity, although it is sizeable
 at these fields:
$g^*\mu_B B=0.11 meV$.Here we assume the bulk giromagnetic factor
for $GaAs$  
 $g^* =-0.44$, but  it could be lower\cite{dobers,nota}.
In any case, at temperatures below $300 mK$ the spin degeneracy
is  lifted. 
According to fig.\ref{fig}, the energy  difference  between 
the GS's when  three 
and two electrons are  on  the dot 
 is $U=^3E-^2E= 12.5 meV$   at the S-T crossing, which is 
 about $1/3$ higher than  the value $E_c + \hbar 
\omega_0  = 3. meV+ 5.55meV$
predicted by a constant interaction model. The $N=3$ particle state
is a doublet ($S=1/2$ )  $^3 D$ with  $M=1$. Higher magnetic fields
produce its  crossing with the doublet with $M=2$, while, increasing
$B_{\perp}$ further, the  quadruplet state ($Q: 
\:\:S=3/2$) becomes lower in energy. 

 The order of magnitude 
of the hybridization of the dot with the contacts $\Delta$ is fixed by 
the requirement that neighbouring CB peaks in the conductance do not 
overlap\cite{cronenwett}. Their separation is $\sim \hbar \omega_0$. 
For values of $\Delta \sim (0.1 -0.2) \hbar \omega _0$, $T_K$ (given
below)  is of tenths of $mK$.
 We have  neglected  the excited
states of the two particle dot in the description. This is 
allowed because 
$T_K$ is much less than the spin splitting of energy levels.
The next available state in the spectrum of the isolated dot with $N=2$ is 
the triplet state with $S_z =0$ ( $^2T_0$). 
Because it is non degenerate, no Kondo-like coupling  is possible. 
This state can only hybridize 
with conduction electrons via cotunneling processes which, in the elastic 
case, give rise to currents that are linear in the biasing voltage  
$V_{sd}$   and of the order of $\sim \left (\Delta^2/U\right )V_{sd}$.

We conclude that 
 the problem is equivalent to a non degenerate Anderson model 
in absence of magnetic field.
 At $B_{\perp}=B^*$,
the $N=2$ particle dot can be in the $^2S_0$ or $^2T_1$ state,
but addition of one more  electron requires an extra energy 
$U$. Furthermore is $U>>\Delta$. 
 The results are expected to be quite 
robust with respect to small changes of  the magnetic field away from 
$B^*$,  especially for  $B_{\perp}< B^*$  because the reference field 
$H_0 = (U \Delta )^{\met }/\mu_B$
is of the order of $Tesla$. For  $B_{\perp} > B^*$, instead,
  the state $^2T_1$ will be eventually favoured in energy.

A small a.c. magnetic field $|B_{\parallel}|
<< |\delta B|  \sim T_K/\mu_B $  ( with 
$\delta B = B_{\perp} - B^* < 0 $) 
 could  induce transitions
between the two degenerate Kondo GS 's, with Rabi frequency $\hbar \omega_R =
M_d |\delta B|$. 
 The quantity $M_d /2 \approx
\met g^*\mu_B \left [ 1 + \delta B /(\pi T_K)\right ]$ is the expected 
magnetic moment of the Dot with $N=2$, in the Kondo state.
 The measurement, if feasible, should be performed just off $ B_{\perp}
=B^*$ for $\delta B < 0$, not to match the energy separation of the 
two levels  $^2T_1$ and $^2T_0$.  

In conclusion, we have discussed the possibility that  the total 
spin S of a dot in  CB  with N=2 is strongly coupled 
 to  the spins of the  electrons  in the leads. This  kind of  
mesoscopic Kondo effect    would give rise to fractionalization 
of  S in the dot, i.e. to a sort of 'spinon box'.

 \vspace*{1cm}
The authors acknowledge useful discussions with B.Altshuler, G.Falci,
L.Glazman, F.Hekking, B.Jouault, B.Kramer,
G.Morandi, H.Sch\"oller, J.Schmid and J.Weis.
Work supported by INFM (Pra97-QTMD )
and by EEC with TMR project, contract FMRX-CT98-0180.


\begin{thebibliography}{99}
\bibitem{nota} note that with  $\hbar \omega_d =8 meV$ and  $E_c =
6meV$ used in ref\cite{faini} the S-T crossing takes place  at $11 Tesla$  
with a spin splitting of $g^*\mu_B B=0.27 meV$.
\end{thebibliography}
\end{document}